\documentclass{llncs}

\usepackage{graphicx}
\usepackage{url}
\usepackage{lscape}
\usepackage{amssymb}
\usepackage{algorithm}
\usepackage{algorithmic}
\usepackage{txfonts}

\DeclareMathAlphabet{\mathtt}{OT1}{pcr}{m}{n}
\SetMathAlphabet{\mathtt}{bold}{OT1}{pcr}{b}{n}

\newcommand{\dataset}{{\cal D}}

\newcommand{\bvec}[1]{\mbox{\boldmath $#1$}}
\newcommand{\sbvec}[1]{\mbox{\scriptsize\boldmath $#1$}}
\newcommand{\fbvec}[1]{\mbox{\footnotesize\boldmath $#1$}}
\newcommand{\vx}{\bvec{x}}
\newcommand{\vy}{\bvec{y}}
\newcommand{\vz}{\bvec{z}}
\newcommand{\svx}{\sbvec{x}}

\newcommand{\svz}{\sbvec{z}}

\newcommand{\fvz}{\fbvec{z}}
\newcommand{\pcx}{p(c\mid\vx)}

\newcommand{\px}{p(\vx)}

\newcommand{\pc}{p(c)}
\newcommand{\pnc}{p(\neg c)}
\newcommand{\pxc}{p(\vx\mid c)}
\newcommand{\pxnc}{p(\vx\mid\neg c)}

\newcommand{\fscore}{{\rm F}}
\newcommand{\true}{\mathit{True}}

\newcommand{\minsup}{\sigma_{\rm{min}}}
\newcommand{\ub}[1]{\overline{#1}}
\newcommand{\sym}[1]{\mathsf{#1}}
\newcommand{\update}{U}
\newcommand{\ptab}{L}
\newcommand{\ccell}[1]{\multicolumn{1}{|c|}{#1}}
\newcommand{\subsumes}{\succ}
\newcommand{\subsumeseq}{\succeq}
\newcommand{\tuple}[1]{\langle{#1}\rangle}
\newcommand{\symitem}[2]{\tuple{#1,#2}}
\newcommand{\numitem}[3]{\tuple{#1,[#2,#3)}}
\newcommand{\cutpoints}{\Delta}
\newcommand{\headertab}{H}
\newcommand{\gridsize}{\varepsilon}

\begin{document}

\begin{center}
\setcounter{footnote}{1}
{\Large Towards Efficient Discriminative Pattern Mining in Hybrid Domains}${}^\thefootnote$
\par
\vspace{1.5em}
\par
{\large Yoshitaka Kameya}\\
Department of Information Engineering, Meijo University\\
\texttt{ykameya@meijo-u.ac.jp}
\end{center}

\footnotetext{This paper is an English version of the paper originally presented in the 17th Forum on Information Technology (FIT 2018), a Japanese domestic conference held during September 19--21, 2018.}

\begin{quote}
\textbf{Abstract.}
Discriminative pattern mining is a data mining task in which we find patterns that distinguish transactions in the class of interest from those in other classes,
and is also called emerging pattern mining or subgroup discovery.
One practical problem in discriminative pattern mining is how to handle numeric values in the input dataset.
In this paper, we propose an algorithm for discriminative pattern mining that
can deal with a transactional dataset in a hybrid domain, i.e.\ the one that includes both symbolic and numeric values.
We also show the execution results of a prototype implementation of the proposed algorithm for two standard benchmark datasets.
\end{quote}

\section{Introduction}
\label{sec:intro}

Discriminative pattern mining is a data mining task in which we find patterns that distinguish transactions in the class of interest from those in other classes,
and is also called emerging pattern mining or subgroup discovery~\cite{Dong12,KraljNovak09}.
Based on discriminative patterns, we do not only obtain some insights from the dataset,
but also build highly reliable classifiers, which are often called associative classifiers~\cite{Thabtah07}.

\begin{figure}[b]
  \centerline{
    \includegraphics[width=.4\hsize]{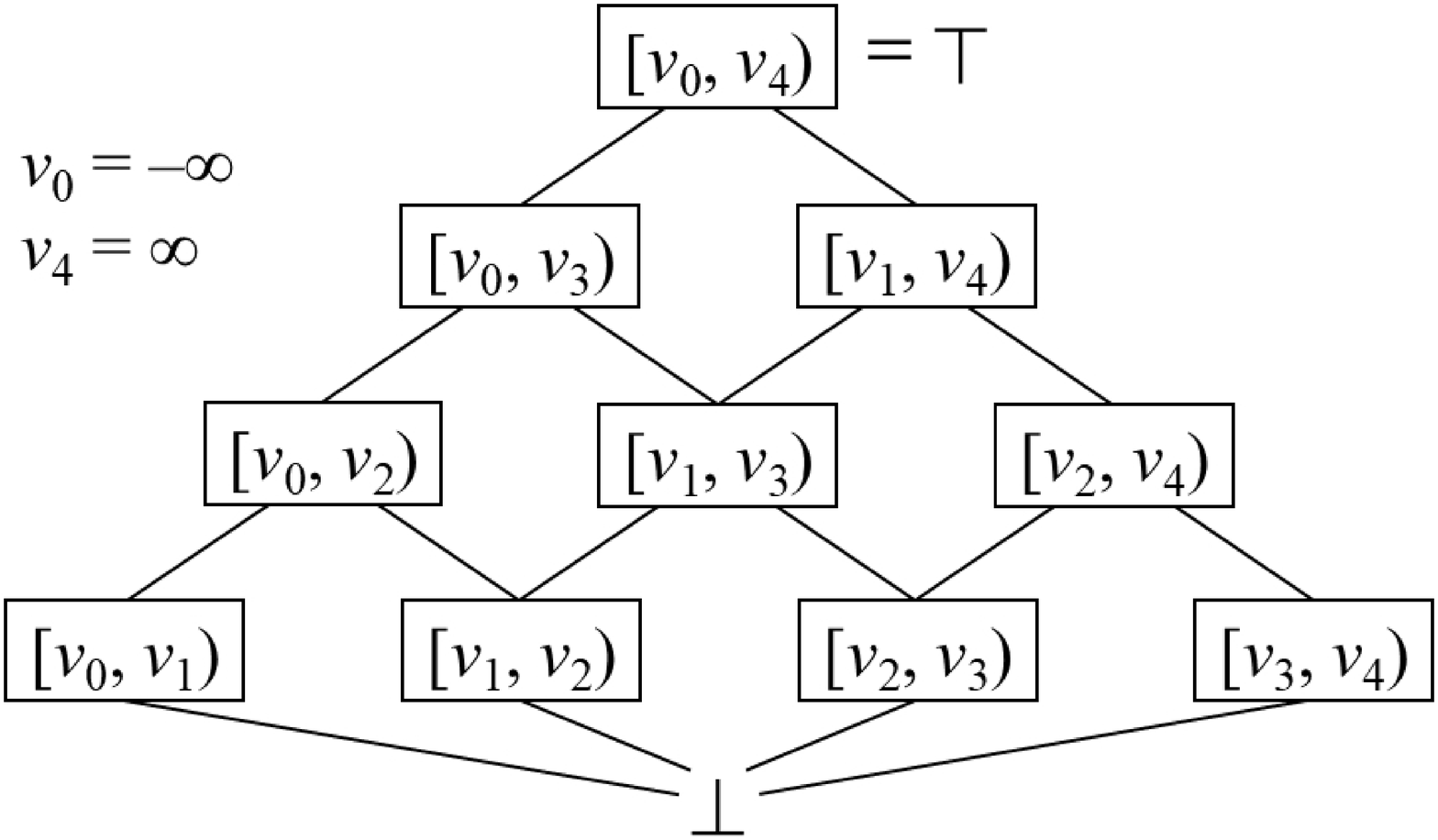}
  }
  \caption{A concept lattice over intervals, where the number $n$ of base intervals is four．}
  \label{fig:lattice}
\end{figure}

One practical problem in discriminative pattern mining is how to handle numeric values in the input dataset.
Of course, since class labels are assumed to be available in discriminative pattern mining,
one may employ some supervised discretization method such as the one proposed by Fayyad and Irani~\cite{Fayyad93}.
However, at present, such a method only works in an attribute-wise manner, and hence may result in an inappropriate discretization.
In addition, let us consider an interval such as $u\le A< v$ as an item $\numitem{A}{u}{v}$,
where $A$ is a numeric attribute, and $u$ and $v$ are the left boundary and the right boundary of the interval, respectively.
Then, such items form a concept lattice, as illustrated Figure~\ref{fig:lattice}.
In the lattice, the intervals $[v_i,v_{i+1})$ at the lowest level except $\bot$ are called the base intervals,
and each raw value in the dataset falls into some base interval.
On the other hand, we consider several upper intervals that subsume base intervals at the higher level in the lattice.
Obviously, we have numerous upper intervals into which each numeric value in the dataset can fall,
and hence the search space for discriminative patterns can be overwhelmingly huge,
where a pattern is a combination of base and upper intervals.
Therefore we can say that, without a sophisticated algorithm under some appropriate constraints over patterns,
it is infeasible to enumerate all discriminative patterns within a practical amount of time.
In some earlier works~\cite{Brin03,Fukuda96}, authors sped up the search by introducing an upper limit in the number of numeric attributes in the dataset.

In this paper, we propose an algorithm for discriminative pattern mining that
can deal with a transactional dataset in a hybrid domain, i.e.\ the one that includes both symbolic and numeric values.
This algorithm is built on top of FP-growth~\cite{Han00}, a standard algorithm for frequent pattern mining,
enhanced with branch-and-bound pruning.
FP-growth stores the transactional dataset into a data structure called an FP-tree,
and conducts a depth-first search with recursive construction of conditional FP-trees.
Since these conditional FP-trees get smaller at deeper places in the enumeration tree,
the search works efficiently despite the vastness of the search space.
Furthermore, FP-trees are designed to be accessed efficiently,
inheriting advantages from both horizontal and vertical layouts.
The proposed algorithm extends FP-trees in order to handle numeric values,
respecting these desirable properties of FP-trees.

Table~\ref{tab:iris} shows discriminative patterns $\vx$ mined by the proposed algorithm
for each class $c$ from the iris dataset provided in the UCI Machine Learning Repository.\footnote{
  \url{http://archive.ics.uci.edu/ml/datasets/Iris}
}
For readability, the interval items in the output patterns are translated into inequality expressions.
In our default setting, the quality of each pattern $\vx$ for class $c$ is measured by F-score, which is denoted as $\fscore_c(\vx)$.
Note here that the patterns in Table~\ref{tab:iris} are considerably long,
presumably because we adopt the closedness constraint,
under which we just perform the least generalization in creating patterns.
Also it is remarkable that
we can have multiple patterns with different balances between confidence (or precision) $\pcx$ and
positive support (or recall) $\pxc$ for classes $\mathsf{versicolor}$ and $\mathsf{virginica}$.
In addition, the search strategy called exhaustive covering guarantees that every positive transactions in the dataset is covered by at least one pattern in Table~\ref{tab:iris}.

The remainder of this paper is structured as follows.
First, in Section~\ref{sec:preliminaries}, we introduce some notations and related concepts, and describe the background of the work.
Section~\ref{sec:proposal} then describes the details of the proposed algorithm.
The results produced by our prototype implementation are reported in Section~\ref{sec:exp},
and finally Section~\ref{sec:conclusion} concludes the paper.

\begin{table}[t]
\caption{Discriminative patterns mined by the proposed algorithm from the iris dataset.}
\label{tab:iris}
\centerline
{\footnotesize
 \def\arraystretch{1.2}
 \tabcolsep=3pt
    \begin{tabular}{|c|ccc|l|}
      \hline
      Class $c$&$\pcx$&$\pxc$&$\fscore_c(\vx)$&\ccell{Pattern $\vx$}\\
      \hline
      $\mathsf{setosa}$&1.000&1.000&1.000&\{$\sym{petal\_len}<2.45$, $\sym{petal\_wid}<0.8$, $\sym{sepal\_len}<5.85$, $2.25\le\sym{sepal\_wid}$\}\\
      \hline
      $\mathsf{versicolor}$&1.000&0.940&0.969&\{$2.45\le\sym{petal\_len}<4.95$, $0.8\le\sym{petal\_wid}<1.65$, $4.85\le\sym{sepal\_len}<7.05$, $\sym{sepal\_wid}<3.45$\}\\
                &0.942&0.980&0.961&\{$2.45\le\sym{petal\_len}<5.15$, $0.8\le\sym{petal\_wid}<1.75$, $4.85\le\sym{sepal\_len}<7.05$, $\sym{sepal\_wid}<3.45$\}\\
                &0.891&0.980&0.933&\{$2.45\le\sym{petal\_len}<5.05$, $0.8\le\sym{petal\_wid}<1.85$, $4.85\le\sym{sepal\_len}<7.05$, $\sym{sepal\_wid}<3.45$\}\\
      \hline
      $\mathsf{virginica}$&0.958&0.920&0.939&\{$4.45\le\sym{petal\_len}$, $1.65\le\sym{petal\_wid}$, $4.85\le\sym{sepal\_len}$, $2.45\le\sym{sepal\_wid}<3.85$\}\\
               &0.891&0.980&0.933&\{$4.75\le\sym{petal\_len}$, $1.35\le\sym{petal\_wid}$, $5.55\le\sym{sepal\_len}$, $2.10\le\sym{sepal\_wid}<3.85$\}\\
      \hline
    \end{tabular}
}
\end{table}

\section{Preliminaries}
\label{sec:preliminaries}

\subsection{Items}
\label{sec:preliminaries:item}

First of all, let us start with introducing some notations and related concepts.
We first assume that the input dataset $\dataset$ is given in a tabular form,
where each instance is represented as a set of attribute-value pairs,
and each attribute-value pair $\tuple{A,v}$ is comprised of an attribute $A$ and its value $v$.
There are two types of attributes, i.e.\ symbolic attributes and numeric attributes.
If an attribute $A$ is symbolic, its value $v$ is chosen from $A$'s own domain,
and if $A$ is numeric, its value $v$ is some real number.

In the proposed algorithm, we transform in advance the input dataset $\dataset$ in this tabular form into the one in a transactional form,
in which an instance corresponds to a transaction.
More specifically, we consider two types of items, i.e.\ symbolic items and interval items.
A symbolic item in a transaction takes the form $\symitem{A}{v}$, and we can equate it with an attribute-value pair in a instance.
An interval item, on the other hand, takes the form $\numitem{A}{u}{v}$, where $A$ is a numeric attribute and $[u,v)$ is an interval,
and further has two types.
A base interval item $\numitem{A}{v_i}{v_{i+1}}$, where $0\le i\le n$, corresponds to one of $n$ base intervals obtained from $n-1$ cut-points
$\cutpoints_A=\{v_1,v_2,\ldots,v_{n-1}\}$ and two extreme points $v_0=-\infty$ and $v_n=\infty$.
An upper interval item is then defined as $\numitem{A}{v_i}{v_j}$ where $0\le i\le n$, $0\le j\le n$, and $i<j$.\footnote{
  In implementation, it would be easier to use a representation like $\tuple{A,i,j}$, where $i$ and $j$ is the indices of cut-points.
}
Note here that, in the proposed algorithm, we avoid having a pattern contain $\numitem{A}{-\infty}{\infty}$, which just means $\true$ in a logical context.
We replace every attribute-value pair $\tuple{A,v_\mathrm{raw}}$ in an instance in $\dataset$ with a base interval item $\numitem{A}{v_i}{v_{i+1}}$ where $v_i\le v_\mathrm{raw} < v_{i+1}$,
and think of each transaction as a set of symbolic items and base interval items.
An item $\eta$ conventionally dealt with in the literature of pattern mining can be represented as $\symitem{\eta}{\bullet}$, where $\bullet$ is an arbitrary constant.
In our default setting, we {\em adaptively} determine the cut-points $\cutpoints_A$ for each numeric attribute $A$ (the details are described in Section~\ref{sec:proposal:discretize}).

\subsection{Transactions}
\label{sec:preliminaries:transaction}

After the transformation above, the input dataset $\dataset$ of size $N$ in a transactional form is denoted as $\{t_1,t_2,\ldots,t_N\}$.
Here, $t_i$ is the $i$-th transaction $(1\le i\le N)$, which is a set of items.
In addition, we assume that each transaction is associated with one of pre-defined class,
and denote the class with which $t_i$ is associated as $c_i$.
It is straightforward to see that each transaction $t_i$ does not have more than one item having the same attribute.

\subsection{Patterns}
\label{sec:preliminaries:pattern}

A pattern $\vx$ is also a set of items having distinct attributes.
However, it should be noted that a pattern can contain symbolic items, base interval items, and upper interval items,
whereas a transaction can only contain symbolic items and base interval items.
In this paper, we use variables like $x$, $y$, $z$, \ldots\ for referring to items.
Furthermore, a pattern $\vx$ is interchangeably interpreted as a vector $\vx=(x_1,x_2,\ldots,x_n)$,
a set $\vx=\{x_1,x_2,\ldots,x_n\}$, or a conjunction $\vx=(x_1\land x_2\land \ldots \land x_n)$ of items,
depending on the context.
An item $x$ may be read as $\{x\}$, a singleton pattern only containing it.

\subsection{Subsumption}
\label{sec:preliminaries:subsume}

There are partial order relations on generality/specificity among items, and among patterns.
To be concrete, for two interval items $x = \numitem{A}{u}{v}$ and $y = \numitem{A}{u'}{v'}$ where $u\le u'$ and $v'\le v$,
we say that $x$ {\em subsumes} $y$, and write this as $x \subsumeseq y$.
This subsumption relation $x\subsumeseq y$ also holds for two symbolic items $x = \symitem{A}{u}$ and $y = \symitem{A}{v}$ where $u=v$.
We also introduce the subsumption relation among patterns.
That is, for two patterns $\vx$ and $\vy$, $\vx\subsumeseq \vy$ holds
when, for each $x\in\vx$, there exists $y\in\vy$ such that $x\subsumeseq y$.
In addition, for two items $x$ and $y$ (resp.\ two patterns $\vx$ and $\vy$),
we say that $x$ strictly subsumes $y$ (resp.\ $\vx$ strictly subsumes $\vy$) and write this as $x\subsumes y$ (resp.\ $\vx\subsumes\vy$),
if and only if $x\subsumeseq y$ and $x\neq y$ (resp.\ $\vx\subsumeseq\vy$ and $\vx\neq\vy$).
As a special case, we say that a pattern $\vx$ {\em covers} a transaction $t_i$ when $\vx\subsumeseq t_i$.

\subsection{Statistics}
\label{sec:preliminaries:stats}

For the class $c$ of interest, we define $\dataset_c(\vx)=\{t_i\mid c_i=c, \vx\subsumeseq t_i, 1\le i\le N\}$.
Using this definition, a set $\dataset_c$ of the transactions that are associated with class $c$ can be written as $\dataset_c(\emptyset)$.
The classes other than $c$ are jointly referred to as a virtual class denoted by $\neg c$,
and we have $\dataset_{\neg c}=\dataset\setminus\dataset_c$,
$\dataset_{\neg c}(\vx)=\dataset(\vx)\setminus\dataset_c(\vx)$, and so on.
Hereafter the transactions in $\dataset_c$ (resp.\ in $\dataset_{\neg c}$) are called positive (resp.\ negative) transactions.

In the proposed algorithm, we think of the empirical probabilities computed from $\dataset$ as basic statistics.
For example, the probability $\pc$ of class $c$'s occurrence is computed as $|\dataset_c|/N$,
and the joint probability $p(c,\vx)$ of class $c$ and pattern $\vx$ is computed as $|\dataset_c(\vx)|/N$.
Then, from joint probabilities, we compute marginal probabilities (e.g.\ $\px$) or conditional probabilities (e.g.\ $\pxc$).
In this paper, $\pxc$ is called positive support, and $\pxnc$ is called negative support.
Note here that, as long as every transaction in the dataset $\dataset$ is associated with a class,
$\pc$ and $\pnc$ can be handled as constant.

\subsection{Relevance}
\label{sec:preliminaries:score}

Given the class $c$ of interest, the quality of a pattern $\vx$ is measured by a score called relevance, and denoted as $R_c(\vx)$.
Many popular relevance scores are defined as functions of positive support $\pxc$ and negative support $\pxnc$.
For example, F-score $\fscore_c(\vx)=2\pcx\pxc/(\pcx + \pxc)$ is known to be equivalent to the Dice coefficient $2p(\vx,c)/(\pc+\px)$,
and is further transformed as a function of positive support and negative support, i.e.\ $2\pc\pxc/(\pc+\pc\pxc+\pnc\pxnc)$.
For a simple classifier having a single rule $\vx\Rightarrow c$, in ROC analysis, positive support is called TPR (true positive rate),
and negative support is called FPR (false positive rate).
Besides, since we are working for characterizing the class $c$ of interest, we just focus on a pattern $\vx$ satisfying
$\pxc\ge\pxnc$, or equivalently $\pcx\ge\pc$.

Here, if $R_c(\vx)$ monotonically increases w.r.t.\ $\pxc$ and decreases w.r.t.\ $\pxnc$ for any pattern $\vx$ satisfying $\pxc\ge\pxnc$,
then relevance $R_c$ is said to be dual-monotonic~\cite{Kameya13}.
In what follows, we derive the criteria for pruning or deleting redundant patterns, assuming dual-monotonicity of relevance.
In fact, many popular relevance scores such as F-score, $\chi^2$, information gain, support difference are dual-monotonic.
It is remarkable that dual-monotonicity is a relaxation of convexity,
which has been assumed in previous work on discriminative pattern mining~\cite{Morishita00,Zimmermann09}.
For example, F-score does not satisfy convexity.
We may add that dual-monotonicity corresponds to two conditions,
out of Piatetsky-Shapiro's three conditions any relevance score must satisfy~\cite{Furnkranz12,PiatetskyShapiro91}.\footnote{
  The remaining one is that $R_c(\vx)$ must be zero if $\pxc=\pxnc$.
}

\subsection{Top-$k$ Mining with Branch-and-Bound Pruning}
\label{sec:preliminaries:branch-and-bound}

In top-$k$ mining, we wish to find only $k$ patterns of the highest relevance.
Let $\vz$ be the pattern of the $k$-th highest relevance at the moment we visit a pattern $\vx$ in a depth-first search,
where we extend patterns by adding items one by one.
Then, in branch-and-bound pruning~\cite{Kameya12a,Morishita00,Zimmermann09},
we obtain an upper bound $\ub{R}_c(\vx)$ in extending $\vx$,
and prune the subtree under $\vx$ if $\ub{R}_c(\vx)<R_c(\vz)$ holds.
To obtain $\ub{R}_c(\vx)$, we forcedly substitute $\pxnc$ with $0$ in the definition of $R_c(\vx)$.\footnote{
  This operation corresponds to computing the relevance in an optimistic situation where FPR turns to be 0 while TPR does not change by extending $\vx$.
}
Here it is easily seen from dual-monotonicity of $R_c$ that $\ub{R}_c(\vx)$ monotonically increases w.r.t.\ $\pxc$,
and therefore decreases w.r.t.\ the extension of $\vx$.
In other words, $\ub{R}_c$ is anti-monotonic w.r.t.\ set-inclusion among patterns.
Since $R_c(\vx')\le\ub{R}_c(\vx')\le\ub{R}_c(\vx)<R_c(\vz)$ holds for $\vx'$ such that $\vx\subset\vx'$,
we can say that the pruning described above is safe.

Moreover, based on the fact that $\ub{R}_c(\vx)$ monotonically increases w.r.t.\ $\pxc$,
one may solve the pruning condition $\ub{R}_c(\vx)<R_c(\vz)$ for $\pxc$\footnote{
  For many relevance scores, $\ub{R}_c(\vx)<R_c(\vz)$ can be solved analytically,
  but for some scores, such as information gain, it must be solved numerically.
}
in order to obtain an inequality of the form $\pxc<\update_c(\vz)$.
For example, an upper bound of F-score is computed as $\ub{\fscore}_c(\vx)=2\pxc/(1+\pxc)$, and from $\ub{\fscore}_c(\vx)<\fscore_c(\vz)$,
we obtain a pruning condition $\pxc<\fscore_c(\vz)/(2-\fscore_c(\vz))$.
To summarize, our branch-and-bound pruning in top-$k$ mining is performed as follows:
(i) we initialize a pruning threshold $\minsup$, called the minimum support, as $1/\dataset_c$,
(ii) we immediately update $\minsup$ by $\minsup:=\max\{\update_c(\vz), \minsup\}$ when we find a new pattern $\vz$ of the $k$-th highest relevance where $R_c(\vz)>R_c(\vz')$ and $\vz'$ is the previous $k$-th  pattern,
and (iii) for a pattern $\vx$ we are visiting, we prune the subtree below $\vx$ if $\pxc<\minsup$.\footnote{
  In FP-growth or its extension, we shrink conditional FP-trees by deleting nodes whose positive support is lower than $\minsup$.
}
$\update_c(\vz)$ monotonically increases w.r.t.\ $R_c(\vz)$,
so the pruning threshold $\minsup$ is continuously raised during the search.

Minimum support raising~\cite{Han02} was originally introduced in frequent pattern mining,
but can also be applied to discriminative pattern mining under dual-monotonic relevance, as described above.
With minimum support raising, it is not necessary to maintain the upper bound $\ub{R}_c$ itself,
and thus we can simplify our mining algorithm by extending some existing frequent mining algorithm such as FP-growth.

\subsection{Constraints among Patterns}
\label{sec:preliminaries:constraint}

One intricate problem in discriminative pattern mining is redundancy among the output patterns.
For example, when an item $x$ is strongly relevant to the class $c$ of interest,
the patterns including $x$ like $\{x,y\}$, $\{x,z\}$, and $\{x,y,z\}$ also tend to be judged as relevant,
and the top-$k$ list can be occupied by such patterns.
To mitigate this problem, a constraint among patterns is often introduced,
and the patterns that violate the constraint are deleted as redundant.

A well-known example of a constraint among patterns in frequent pattern mining should be the closedness constraint~\cite{Pasquier99}.
In discriminative pattern mining, on the other hand, we often consider the closedness constraint over positive transactions~\cite{Garriga08},
which we hereafter call the closedness-on-the-positives constraint.
Under this constraint, we consider that the patterns covering the same positive transactions form an equivalence class,
and only leave the most specific pattern $\vx$ w.r.t.\ $\subsumeseq$ among the patterns in the equivalence class containing $\vx$.

The best-covering constraint~\cite{Kameya16} says that each output pattern $\vx$ must have some positive transaction covered by $\vx$ with the highest relevance.
This constraint is a generalization of the ``highest confidence covering'' constraint introduced in an associative classifier called HARMONY~\cite{Wang05}, 
to the case of discriminative pattern mining under dual-monotonic relevance.
It is shown in \cite{Kameya16} that the best-covering constraint is tighter than the productivity constraint~\cite{Bayardo00,Kameya12a,Webb07},
which has been used in the literature of discriminative pattern mining.

Under dual-monotonic relevance, a pattern $\vx$ satisfying the closedness-on-the-positives constraint has the highest relevance among the patterns in the equivalence class containing $\vx$~\cite{Garriga08,Soulet04},
and in this sense, the best-covering constraint and the closedness-on-the-positives constraint are basically consistent.
However, there is an exceptional case that two patterns $\vx$ and $\vx'$ such that $\vx\subsumes\vx'$ cover the same positive transactions and have the same relevance score.
In such a case, the best-covering constraint prefers $\vx$, a more general one, whereas the closedness-on-the-positives constraint prefers $\vx'$, a more specific one,
and following \cite{Kameya16},
the proposed algorithm gives priority to the closedness-on-the-positives constraint．

\subsection{Exhaustive Covering}
\label{sec:preliminaries:excover}

Sequential covering is known as a standard method for rule learning~\cite{Friedman99,Furnkranz12,Witten05}.
In sequential covering, we first extract a rule having the class of interest in the consequent,
and then delete all positive instances covered by the rule.\footnote{
  In building associative classifiers, a method called database coverage performs a similar operation for rule reduction~\cite{Thabtah07}.
}
One problem in sequential covering is that deleting positive instances is quite procedural and hence we cannot interpret the extracted rules declaratively.
It should also be noted that deleting positive instances makes the extracted rules unreliable in a statistical sense.
To overcome these problems, Domingos~et~al. proposed a strategy called ``conquering without separating,'' 
in which we learn each rule from the entire dataset~\cite{Domingos94}．

Exhaustive covering~\cite{Kameya16} was proposed in the same spirit of ``conquering without separating.''
Exhaustive covering works under the best-covering constraint, and conducts top-1 pattern mining concurrently for each positive transaction.
Specifically, we first prepare a candidate list $\ptab[t']$ for each positive transaction $t'$ in the entire dataset (i.e.\ $t'\in\dataset_c$).
Then, when we visit a candidate pattern $\vx$ in a depth-first search,
for each positive transaction $t$ covered by $\vx$ (i.e.\ $t\in\dataset_c(\vx)$), we check its candidate list $\ptab[t]$.
If $\ptab[t]$ is empty, we just add $\vx$ into $\ptab[t]$.
Otherwise, we compare the relevance $R_c(\vx)$ of $\vx$ and the relevance $R_c(\vz)$ of a tentative top-1 pattern $\vz$ in $\ptab[t]$.
If $R_c(\vx)>R_c(\vz)$, we add $\vx$ into $\ptab[t]$ after deleting all patterns from $\ptab[t]$.
This means that $\vx$ becomes the only tentative top-1 pattern in $\ptab[t]$.
If $R_c(\vx)=R_c(\vz)$, we just add $\vx$ into $\ptab[t]$ as another tentative top-1 pattern.
If $R_c(\vx)<R_c(\vz)$, we give up adding $\vx$ into $\ptab[t]$.
We iteratively perform the operation above every time visiting a new candidate pattern.
After the search has finished, we return $\bigcup_{t'\in\dataset_c} \ptab[t']$ as the set of output patterns.
During the search, the threshold $\minsup$ in branch-and-bound pruning (Section~\ref{sec:preliminaries:branch-and-bound}) is raised by $\minsup:=\max\{\update_c^\ast(\vz),\minsup\}$,
where $\update_c^\ast(\vz)=\min_{t\in\dataset_c(\svx)}\max_{\svz\in\ptab[t]}\update_c(\vz)$,
and $\max_{\fvz\in\ptab[t]}\update_c(\vz)$ is exceptionally defined as $1/|\dataset_c|$ when $\ptab[t]$ is empty.

The entire search finishes at the moment the top-1 pattern has been determined for every positive transaction.
Top-$k$ mining has an advantage in user-centricity that we only need to specify the number $k$ of patterns to be output,
instead of the minimum support $\minsup$ which is known as a sensitive threshold.
Furthermore, exhaustive covering does not require even $k$.

\subsection{Dynamic Re-ordering}
\label{sec:preliminaries:reorder}

As is seen from the description in Section~\ref{sec:preliminaries:branch-and-bound},
in top-$k$ pattern mining, branch-and-bound pruning will be more effective if patterns of higher relevance are found earlier.
Therefore, it is reasonable to introduce dynamic re-ordering~\cite{Atzmueller09,Kameya17,Webb95} at branches in the depth-first search.
Besides, even in a case that we have to terminate the search after a permitted amount of time,
dynamic re-ordering would enable our mining algorithm to work as an anytime algorithm that can leave patterns of higher relevance in the output.

\section{The Proposed Algorithm}
\label{sec:proposal}

Based on exhaustive covering built on top of FP-growth,
this paper proposes an algorithm for discriminative pattern mining that can deal with a transactional dataset including both symbolic and numeric values.
In the proposed algorithm, we also reduce the redundancy among patterns with a help from the best-covering constraint and the closedness-on-the-positives constraint.
Since the proposed algorithm basically follows the procedure described in the sections from Section~\ref{sec:preliminaries:branch-and-bound} to Section~\ref{sec:preliminaries:excover},
in the sequel, we mainly describe the part newly introduced for dealing with numeric values.

\subsection{Creating Base Interval Items}
\label{sec:proposal:discretize}

As described in Section~\ref{sec:preliminaries:item}, in transforming the input dataset into
a transactional form, we need to determine a set $\cutpoints_A$ of cut-points for converting each value of a numeric attribute $A$ into a base interval item.

First, for each numeric attribute $A$, we collect $A$'s values appearing in the dataset in ascending order,
and consider the mid-points between neighboring values as initial cut-points.
Here we regard all values in $[m\gridsize,(m+1)\gridsize)$ as identical ($m=0,1,2,\ldots$),
where $\gridsize$ is a pre-defined, sufficiently small precision.
Also note that we have one or more values between neighboring initial cut-points. 
Next, when attribute $A$ takes a value $v$ in the $i$-th transaction $t_i$,
we associate $v$ with class $c_i$, and see that there are regions between neighboring initial cut-points,
containing (i) only the values associated with class $c$ (the positive class),
(ii) only the values associated with class $\neg c$ (the negative class),
and (iii) the values mixedly associated with $c$ or $\neg c$.
In the proposed algorithm, we merge two or more consecutive regions satisfying condition (i) to the maximum extent,
and also merge two or more consecutive regions satisfying condition (ii) to the maximum extent.
Finally, $\cutpoints_A$ is a set of the cut-points obtained after this merge operation has been exhausted.

The merge operation above was firstly introduced by Brin~et~al.~\cite{Brin03},
and used later by Grosskreutz~et~al.~\cite{Grosskreutz09}.
In the proposed algorithm, on the other hand, we can justify this merge operation in a general way,
using the notions from dual-monotonicity in relevance and the best-covering constraint.
First, we will justify the merge operation based on condition (i).
Let $[u,v)$ be a merged region which contains only values associated with the positive class.
We also consider other cut-points $v_l$, $t$, and $v_r$ such that $v_l\le u<t<v\le v_r$.
Note here that $t$ is a cut-point located in the middle of the region $[u,v)$.
Then, we have
\begin{eqnarray*}
p(\numitem{A}{v_l}{t}\cup\vx\mid c)&<&p(\numitem{A}{v_l}{v}\cup\vx\mid c),\\
p(\numitem{A}{v_l}{t}\cup\vx\mid\neg c)&=&p(\numitem{A}{v_l}{v}\cup\vx\mid\neg c),\\
p(\numitem{A}{t}{v_r}\cup\vx\mid c)&<&p(\numitem{A}{u}{v_r}\cup\vx\mid c),\\
p(\numitem{A}{t}{v_r}\cup\vx\mid\neg c)&=&p(\numitem{A}{u}{v_r}\cup\vx\mid\neg c)
\end{eqnarray*}
where $\vx$ is an arbitrary pattern.
From dual-monotonicity of relevance and the first two relations above,
we see that $R_c(\numitem{A}{v_l}{t}\cup\vx)<R_c(\numitem{A}{v_l}{v}\cup\vx)$.
Generally speaking, for two patterns $\vy$ and $\vy'$ such that $\vy$ strictly subsumes $\vy'$ and is more relevant than $\vy'$,
$\vy'$ violates the best-covering constraint~\cite{Kameya16}.
So in the case above, a more specific pattern $\numitem{A}{v_l}{t}\cup\vx$ must be excluded under the best-covering constraint.
Similarly, from dual-monotonicity of relevance and the last two relations above,
$\numitem{A}{t}{v_r}\cup\vx$ must be excluded.
Consequently, there is no reason for leaving $t$ as a cut-point.
The merge operation based on condition (ii) is also justified by a similar discussion.

\subsection{Computing Relevance of Interval Items}
\label{sec:proposal:score}

In an extended FP-growth for discriminative pattern mining~\cite{Kameya12a},
we construct a conditional FP-tree for each candidate pattern $\vx$ being visited.
During the process of construction, we count the occurrences of each item $x$ in the positive (resp.\ negative) conditional transactions,
and interpret it as $N_c(x\cup\vx)=|\dataset_c(x\cup\vx)|$ (resp.\ $N_{\neg c}(x\cup\vx)=|\dataset_{\neg c}(x\cup\vx)|$).
Then, from these occurrences, we compute positive support $p(x\cup\vx\mid c)$, negative support $p(x\cup\vx\mid\neg c)$, and relevance $R_c(\vx)$.

Conditional transactions only contain symbolic items and base interval items,
so we compute the positive support and the negative support of upper interval items by dynamic programming over a concept lattice like the one in Fig.~\ref{fig:lattice}.\footnote{
  We can straightforwardly represent such a concept lattice as a triangular matrix implemented in a two-dimensional array.
}
To be more specific, we first consider $\cutpoints_A=\{v_1,v_2,\ldots,v_{n-1}\}$, $v_0=-\infty$, and $v_n=\infty$.
Then, for $d=2,3,\ldots,n-1$, we compute positive support in turn by
\[
  p(\numitem{A}{v_i}{v_{i+d}}\cup\vx\mid c):=p(\numitem{A}{v_i}{v_{i+1}}\cup\vx\mid c)+p(\numitem{A}{v_{i+1}}{v_{i+d}}\cup\vx\mid c),
\]
where $0\le i\le n-d$.
Negative support is also computed in a similar manner.

In addition, we delete every symbolic item $x$ that appears in conditional transactions but has positive support $p(x\cup\vx\mid c)$ lower than $\minsup$.
Such an item is also excluded from the header table associated with the conditional FP-tree (see Section~\ref{sec:proposal:fp-tree} for the details of the header table).
By this operation, we will not visit $x\cup\vx$ in the further search.
We should however note that a base interval item of attribute $A$ can be deleted only when the total sum of the positive support of $A$'s base interval items is lower than $\minsup$.

Having computed the positive support of upper interval items, we conduct a pruning based on the closedness-on-the-positives constraint.
More specifically, let us suppose that $p(\numitem{A}{v_{i}}{v_{j}}\cup\vx\mid c)=p(\numitem{A}{v_{i}}{v_{j-1}}\cup\vx\mid c)$ or $p(\numitem{A}{v_{i}}{v_{j}}\cup\vx\mid c)=p(\numitem{A}{v_{i+1}}{v_j}\cup\vx\mid c)$.
This means that an upper interval item $\vy=\numitem{A}{v_{i}}{v_{j}}\cup\vx$ has the same positive support as that of its specialization.
Now $\vy$ is said to violate the closedness-on-the-positives constraint, and therefore is excluded from the header table.

\subsection{Extending FP-Trees for Handling Interval Items}
\label{sec:proposal:fp-tree}

In the original FP-growth, while constructing a conditional FP-tree, we record each item $x$ appearing in conditional transactions into the header table $\headertab$.
At the same time, a list connecting the nodes indicating $x$ inside the conditional FP-tree is created and recorded into $\headertab$.
Hereafter the list is referred to as $\headertab[x]$.

Since, as mentioned above, conditional transactions only contain symbolic items and base interval items,
it is additionally required to record each upper interval item $y$ into the header table together with a list connecting the nodes indicating $y$.
Let us consider a numeric attribute $A$ such that $\cutpoints_A=\{v_1,v_2,\ldots,v_{n-1}\}$, $v_0=-\infty$, and $v_n=\infty$.
Then, at the moment we insert a node $w$ which indicates a base interval item $x=\numitem{A}{v_i}{v_{i+1}}$ into a conditional FP-tree,
we consider an upper interval item $y=\numitem{A}{v_j}{v_{j'}}$ that strictly subsumes $x$,
where $0\le j\le i$ and $i+1\le j'\le n$, and $y\neq\numitem{A}{-\infty}{\infty}$.
For an upper interval item $y$ that satisfies $p(y\cup\vx\mid c)\ge\minsup$ and the closedness-on-the-positives constraint,
we update the contents of the header table $\headertab$.
That is, if $y$ has not been recorded in $H$ yet,
we record $y$ into $\headertab$ and let $w$ be the only element of a new list $\headertab[y]$.
Otherwise, we just add $w$ into the existing list $\headertab[y]$.

One may find that it is tedious to test whether $p(y\cup\vx\mid c)\ge\minsup$
for every upper interval item $y=\numitem{A}{v_j}{v_{j'}}$ that subsumes a base interval item $x=\numitem{A}{v_i}{v_{i+1}}$ appearing in conditional transactions.
In order to narrow down the upper interval items to be tested, we exploit the structure of a concept lattice.
That is, we traverse the concept lattice from higher levels to lower levels,
and once we find an upper interval item $y$ such that $p(y\cup\vx\mid c)<\minsup$,
we will not test the upper interval items subsumed by $y$,
by dynamically modifying the ranges $j$ and $j'$ move.

\subsection{Dynamic Merge of Base Intervals}
\label{sec:proposal:merge}

As described in Section~\ref{sec:proposal:discretize},
we create base interval items by merging the consecutive regions that contain only
the values associated with the positive class, or the values with the negative class.
It is further observed that, in conditional FP-trees constructed during the search,
base interval items tend to appear only in positive conditional transactions,
or only in negative conditional transactions.
This is because a set of conditional transactions is necessarily smaller than a set of the original transactions.
and the imbalance between positivity and negativity in each base interval item's appearance gets significant.

Based on the observation above,
we dynamically merge two or more consecutive base interval items appearing only in positive (or negative) conditional transactions,
into a corresponding upper interval item.
Concept lattices are also re-organized accordingly.
This operation is surely costly to some extent,
but there is a computational merit that it accelerates the dynamic-programming-based computation of relevance over a concept lattice (Section~\ref{sec:proposal:score}),
and reduces the number of branches in the depth-first search.

\subsection{Exploiting the Closedness Constraint}
\label{sec:proposal:closed}

Section~\ref{sec:preliminaries:constraint} described that the proposed algorithm works under two constraints among patterns,
i.e.\ the best-covering constraint and the closedness-on-the-positives constraint.
An existing algorithm that follows the best-covering constraint~\cite{Kameya16} performs the closure operation originally proposed in LCM~\cite{Uno04},
a well-known algorithm for frequent pattern mining,
in order to traverse only over the patterns satisfying the closed-on-the-positives constraint.
However, currently we do not use the closure operation.
Instead, during the search, we just store possibly more than one pattern of the highest relevance into each candidate list $\ptab[\cdot]$,
and after the search, we test such patterns firstly by the closed-on-the-positives constraint, and secondly by the best-covering constraint.
Finally we delete the patterns violating these constraints.
The reasons why we do not use the closure operation currently are the powerfulness of pruning based on the best-covering constraint and the ease of implementation.
It should be studied in future how efficiently the closure operation can be performed in our task,
and how effectively the search space can be shrunk by the closure operation,
and so on.

\subsection{Notes on Dynamic Re-ordering}
\label{sec:proposal:reorder}

In dynamic re-ordering (Section~\ref{sec:preliminaries:reorder}),
we can re-order the items to be added at a branch of the depth-first search in descending order w.r.t.\ relevance,
regardless of the type of items (symbolic items, base interval items, and upper interval items).\footnote{
  Contrastingly, in Grosskreutz~et~al.'s method~\cite{Grosskreutz09}, it is required to add symbolic items first.
}
On the other hand, it should be noted that the order of items in a conditional transaction does not have to coincide with the order of items at branches,
and that the items of different attributes must not be arranged in an interleaved manner.

\section{Experiments}
\label{sec:exp}

In this section, we report the results of our experiment with a prototype implementation of the proposed algorithm written in Java.
The input dataset is the german credit dataset from UCI Machine Learning Repository.\footnote{
  \url{https://archive.ics.uci.edu/ml/datasets/Statlog+(German+Credit+Data)}
}
This dataset classifies customers as good or bad credit risks,
and contains 1,000 instances and 20 attributes (7 numeric ones and 13 symbolic ones).
It seems that some of symbolic attributes were originally numeric and have been discretized manually.

In the experiment, we run the prototype implementation to obtain discriminative patterns for the $\sym{good}$ class and the $\sym{bad}$ class.
The processor we used is Core i7 3820 (3.6GHz).
For the $\sym{good}$ class, we visited 28,491,232 candidate patterns and it took 686 seconds (nearly 12 minutes) until the search had finished.
For the $\sym{bad}$ class, we visited 2,100,728,918 candidate patterns and it took 24,707 seconds (nearly 7 hours).

Table~\ref{tab:credit-g} shows the discriminative patterns mined for each class.
We can see from these patterns that a customer with a smaller credit amount and a short duration can be classified as a good customer.
In contrast, it seems not straightforward to judge a customer as a bad customer, and additionally we need to take into account age, saving status, being a foreign worker or not, and so on.

\section{Concluding Remarks}
\label{sec:conclusion}

In this paper, we proposed an algorithm for discriminative pattern mining that can deal with a transactional dataset including both symbolic and numeric values,
based on exhaustive covering built on top of FP-growth.
We also showed the execution results of a prototype implementation of the proposed algorithm for two standard benchmark datasets
i.e.\ the iris dataset and the german credit dataset.

In future, we would like further to investigate the characteristics of the proposed algorithm using other benchmark datasets.
Now we comprehend that the number of base intervals, which are adaptively determined from an input dataset, severely affects the runtime.
So, as a remedy, it would be promising to introduce an attribute-wise, but non-greedy supervised discretization method based on a histogram construction method~\cite{Kontokanen07}.
We are also planning to develop associative classifiers that exploit the discriminative patterns obtained by the proposed algorithm.

\begin{landscape}
\begin{table}[b]
\caption{Discriminative patterns mined by the proposed algorithm from the german credit dataset.}
\label{tab:credit-g}
\rightline{\footnotesize
 \def\arraystretch{1.2}
 \tabcolsep=3pt
    \begin{tabular}{|c|ccc|l|}
      \hline
      Class $c$&$\pcx$&$\pxc$&$\fscore_c(\vx)$&\ccell{Pattern $\vx$}\\
      \hline
      $\mathsf{good}$
          &0.714&0.987&0.829&\{$\sym{credit\_amount}<10920$, $\sym{duration}<66$\}\\
          &0.713&0.989&0.828&\{$\sym{credit\_amount}<11190$, $\sym{duration}<66$\}\\
          &0.711&0.990&0.827&\{$\sym{credit\_amount}<11790$, $\sym{duration}<66$\}\\
          &0.709&0.993&0.827&\{$\sym{credit\_amount}<12300$, $\sym{duration}<66$\}\\
          &0.706&0.997&0.827&\{$\sym{credit\_amount}<14250$, $\sym{duration}<66$\}\\
          &0.702&1.000&0.825&\{$\sym{credit\_amount}<15900$, $\sym{duration}<66$\}\\
      \hline
      $\mathsf{bad}$
          &0.411&0.647&0.503&\{$19.5\le\sym{age}<61.5$, $430.5\le\sym{credit\_amount}$, $11.5\le\sym{duration}<66$, $\sym{exist\_cred}<3.5$, $\sym{savings\_status}=\sym{less\_than\_100}$\}\\
          &0.407&0.657&0.503&\{$19.5\le\sym{age}<66.5$, $430.5\le\sym{credit\_amount}$, $11.5\le\sym{duration}<66$, $\sym{exist\_cred}<3.5$, $\sym{savings\_status}=\sym{less\_than\_100}$\}\\
          &0.401&0.673&0.502&\{$\sym{age}<61.5$, $430.5\le\sym{credit\_amount}$, $8.5\le\sym{duration}<66$, $\sym{exist\_cred}<2.5$, $\sym{savings\_status}=\sym{less\_than\_100}$\}\\
          &0.403&0.663&0.501&\{$\sym{age}<61.5$, $430.5\le\sym{credit\_amount}$, $7.5\le\sym{duration}<66$, $\sym{exist\_cred}<2.5$, $\sym{foreign\_worker}=\sym{yes}$, $\sym{savings\_status}=\sym{less\_than\_100}$\}\\
          &0.395&0.683&0.501&\{$\sym{age}<74.5$, $605\le\sym{credit\_amount}$, $8.5\le\sym{duration}<66$, $\sym{exist\_cred}<2.5$, $\sym{savings\_status}=\sym{less\_than\_100}$\}\\
          &0.406&0.630&0.494&\{$\sym{age}<61.5$, $430.5\le\sym{credit\_amount}$, $5.5\le\sym{duration}<66$, $\sym{exist\_cred}<2.5$, $\sym{foreign\_worker}=\sym{yes}$, $1.5\le\sym{install\_commit}$, $\sym{savings\_status}=\sym{less\_than\_100}$\}\\
          &0.388&0.680&0.494&\{$\sym{age}<61.5$, $430.5\le\sym{credit\_amount}$, $5.5\le\sym{duration}<66$, $\sym{exist\_cred}<2.5$, $\sym{foreign\_worker}=\sym{yes}$, $\sym{savings\_status}=\sym{less\_than\_100}$\}\\
          &0.382&0.697&0.494&\{$\sym{age}<74.5$, $430.5\le\sym{credit\_amount}$, $5.5\le\sym{duration}<66$, $\sym{exist\_cred}<2.5$, $\sym{foreign\_worker}=\sym{yes}$, $\sym{savings\_status}=\sym{less\_than\_100}$\}\\
          &0.341&0.883&0.493&\{$19.5\le\sym{age}<61.5$, $430.5\le\sym{credit\_amount}$, $11.5\le\sym{duration}$, $\sym{foreign\_worker}=\sym{yes}$\}\\
          &0.338&0.900&0.492&\{$19.5\le\sym{age}<69$, $430.5\le\sym{credit\_amount}$, $11.5\le\sym{duration}$, $\sym{foreign\_worker}=\sym{yes}$\}\\
          &0.332&0.933&0.490&\{$\sym{age}<61.5$, $430.5\le\sym{credit\_amount}$, $8.5\le\sym{duration}$, $\sym{foreign\_worker}=\sym{yes}$\}\\
          &0.318&0.963&0.478&\{$\sym{age}<61.5$, $430.5\le\sym{credit\_amount}$, $5.5\le\sym{duration}$, $\sym{foreign\_worker}=\sym{yes}$\}\\
      \hline
    \end{tabular}
}
\end{table}
\end{landscape}

\end{document}